\documentstyle[aas2pp4]{article}

\def\Msun{\ifmmode M_{\odot} \else $M_{\odot}$\fi}
\def\Lsun{\ifmmode L_{\odot} \else $L_{\odot}$\fi}
\def\eg{{\it e.g.,\ }}

\def\etal{{et al.~}}

\lefthead{Chatzichristou}
\righthead{IR-Warm Seyfert Galaxies}

\begin{document}

\title{Multicolour Optical Imaging of IR-Warm Seyfert Galaxies.\\
    IV. Surface Photometry: Colour Distributions}

\author{Eleni T. Chatzichristou}
\affil{Leiden Observatory, P.O. Box 9513, 2300 RA Leiden, The Netherlands}

\affil{NASA/Goddard Space Flight Center, Code 681, Greenbelt, MD 20771}



\begin{abstract}
This paper is the fourth in a series, studying the optical properties of a
sample of mid-IR Warm Seyfert galaxies and of a control sample of mid-IR cold
galaxies. The present paper is devoted to the analysis of the colour distributions
characterizing the host galaxies. The Warm Seyfert 1 and 2 galaxies show
opposite colour gradients and their colour profiles are depicting age 
and dust effects within single-burst, solar metallicity models. In particular,
we find ample evidence for the occurrence of strong star formation in the 
Seyfert 2 disks: their colour and emission line two-dimensional maps suggest 
dust extinction associated with on-going star formation in spiral and tidal 
features; their colour profiles show starbursts of 0.5-1 Gyr or younger, 
superposed on the older underlying galaxy population. Most of these properties
are shared with the Cold galaxies, while the Warm Seyfert 1s show mostly older
stellar populations and, in only a few cases, evidence for circumnuclear star 
formation.
\end{abstract}




%

\section{Introduction}

The present paper is devoted to the analysis of the optical colour 
distributions in a subsample of 54 mid-IR Warm Seyferts selected from the 
original sample 
of IR-warm IRAS sources of De Grijp \etal (1987 and 1992). This is compared to
a control sample of 16 mid-IR cold IRAS galaxies, selected to span 
similar redshift and luminosity ranges as the Warm sample.
In \cite{paper1} (hereafter Paper I) we presented our optical imaging data.
In \cite{paper2} (hereafter Paper II) we discussed and intercompared the
optical properties of these samples, resulting from aperture photometry.
In \cite{paper3} (hereafter Paper III) we parametrized the bulge and disk
components of their host galaxies.
In the present paper, we present and discuss extensively the colour 
distributions and their implications for the stellar and dust content of 
IR-warm Seyferts, in terms of 2D colour maps, radial colour gradients and 
colour-colour plots, in Sections 2, 3 and 4, respectively. In Section 5 we
summarize our conclusions. We also refer the reader to the 
Appendix of Paper III. where  we presented radial colour profiles, 
colour-colour plots and 2D colour and emission line maps, together with 
a brief description, for each individual galaxy.

\section{Colour and Emission-Line Maps}

\subsection{Presentation of Data}

In the Appendix of Paper III we presented two dimensional colour maps for most
of our sample objects. These were constructed after: (i) correcting the 
individual band images for geometrical distortion and aligning them, (ii) 
degrading the images to match the resolution of those with the worst seeing 
(the seeing for each image, as measured from field stars, is given in
Tables 4 and 5 of Paper I), (iii) applying a ``reasonable'' noise cutoff 
($\sim$2$\sigma$) to all individual images before constructing the colour 
maps (iv) using the appropriate calibration formulas (the same used for the 
aperture magnitudes in Paper II) for each observing session, (v) convolving 
with a circular Gaussian function of $\sigma$=0.5 pix, to produce the smooth 
images shown in the Appendix.

The grey scales in most colour images range from 0-2 mags, except for the 
Seyfert 1s IRAS 01378-2230 and 04339-1028, the Seyfert 2 IRAS 03230-5800 and
the Cold galaxy IRAS 23128-5919. These are objects with no available
photometric calibration, for which surface magnitudes and colours have been 
estimated using a reasonable assumption for the mean galaxy colours (based on the 
other sample objects), but we do not include these estimated quantities in the
quantitative analysis presented below. For the objects for which no colour 
maps are shown, this is either because they were too noisy (original images 
shallow) or because we have data only in one band.

When H$\alpha$ narrow band images are available, they are shown next to the 
colour maps. The H$\alpha$ images were produced by: (i) subtracting the 
background and aligning them with the broad band images, (ii) using the 
latter to estimate and subtract the continuum. The best result is judged by
the smooth residuals in the resulting (continuum subtracted) image, everywhere
in the galaxy except for the line emitting regions. The narrow band images are
not flux calibrated, thus we used them to make qualitative morphological 
comparisons with the colour maps but cannot extract H$\alpha$ luminosities.

\subsection{Results}

We summarize here the qualitative conclusions drawn from inspection of the 
two-dimensional colour and emission line maps presented in the Appendix of
Paper III. 

Seyfert 1s: Their colour maps are mostly featureless. Almost all of them have 
blue central regions and colour gradients that become more positive (redder) 
outwards. Exceptions are (i) IRAS 14557-2830, a strongly interacting system 
with patchy red central regions and (ii) IRAS 23016+2221 and 04339-1028, both 
showing structure in their colour distributions associated with individual 
morphological peculiarities. We have only two Seyfert 1s with available 
H$\alpha$ maps: the line emission follows roughly the continuum light emission
and is mainly centrally concentrated and/or delineates knotty spiral arms.

Seyfert 2s: Most galaxies have redder central regions with negative (bluer)
colour gradients outwards. The colour distributions are more complex than in 
the case of Seyfert 1s: (a) In many cases narrow red features are seen as
(i) one-sided red strips (\eg IRAS 04507+0358 and 20481-5715), (ii) lanes 
associated with inter-arm regions (\eg IRAS 23254+0830), edge-on disks 
(\eg IRAS 11298+5313E), or central bars (\eg IRAS 02580-1136 and 03059-2309). 
These features are most probably due to dust extinction or, when associated 
with a bar, to an older stellar population. (b) Blue bright emission 
knots are observed, associated with spiral arms (\eg IRAS 02580-1136,
03059-2309, 23254+0830) or ring structures (\eg IRAS 03202-5150). These 
blue features are likely to be star forming regions, as seen by comparison of
the colour and  H$\alpha$ emission line maps. (c) There are two double 
nucleus merger products in our Seyfert 2 sample (IRAS 13536+1836 and
19254-7245) that show remarkably similar colour and line emission 
distributions: (i) the brightest of the two nuclei is also much redder; (ii)
the line emission is centered on that nucleus and shows a spike-like emission 
feature perpendicular to the line connecting the two nuclei; (iii) blue knotty
arms are seen in both these galaxies, probably associated with intense star 
formation. The detailed spectroscopic analysis and mapping of IRAS 13536+1836 extended emission-line region (\cite{eleni95}) have shown that the 
brighter/redder is also the Seyfert nucleus and that the gas is 
anisotropically ionized by the AGN. We suspect that this must also be the case
for IRAS 19254-7245, given the similarity of features between the two objects.

(3) Cold galaxies: The vast majority of the Cold sample galaxies are strongly 
interacting systems, with complex morphologies and colour distributions: 
(i) Blue regions are associated with circumnuclear rings, (knotty) 
spiral arms, tidal loops and disk distortions, most probably indicating 
strong star formation events. (ii) Red features are irregular and patchy, or 
appear as one-sided lanes, associated with bars and edge-on disks. They most
probably are due to a combination of dust effects and redder stellar 
populations, as in the case of Seyfert 2s. (iii) There is one double nucleus 
merger in our Cold sample (IRAS 23128-5919) that shows strikingly similar 
characteristics to the Warm Seyfert 2 mergers: the brighter and redder of 
the two nuclei in continuum light is also brighter in H$\alpha$ light and the 
line emission shows a spike perpendicular to the line connecting the two 
nuclei. It is interesting that this object is spectroscopically classified 
as a Seyfert 2.
  
We have observed an additional double nucleus system, the Seyfert 1 IRAS
19580-1818, not presented here because of its low S/N broad and narrow band
images (a broad-band contour map is shown in Paper I). Its continuum 
and emission line characteristics are again strikingly similar to the ones 
described above for the Seyfert 2 and Cold merger systems.

\section{Radial Colour Profiles and Colour Gradients}

In the Appendix of Paper III, we have presented the (azimuthally averaged) 
radial colour profiles of our objects, constructed using surface magnitudes  
(corrected for Galactic absorption and K-correction). Consequently, labels 
such as $(B-V)$ signify $\mu_{B}-\mu_{V}$. The error bars are derived from the
surface magnitude errors and are representative of the photometric 
uncertainties, except in the outer regions where uncertainties in the sky 
subtraction dominate (for details see ~\cite{thesis}). To avoid the noisy outer regions, we plotted radial 
colour profiles out to $\mu_{B}$=25 mag arcsec$^{-2}$. The horizontal 
axis is in linear scale and units of kpc and represents the galaxy's 
semi-major axis lengths. The inner cut-off in the profiles is applied to avoid
seeing effects and is approximately equal to the seeing radius, as measured on
the field stars. In what follows we will be discussing these data in terms of 
surface colour gradients and colour-colour plots. But first, we will
summarize our results from the aperture colour gradients.

\subsection{Aperture Colour Gradients}

\begin{figure}
\epsscale{1.}
\plotone{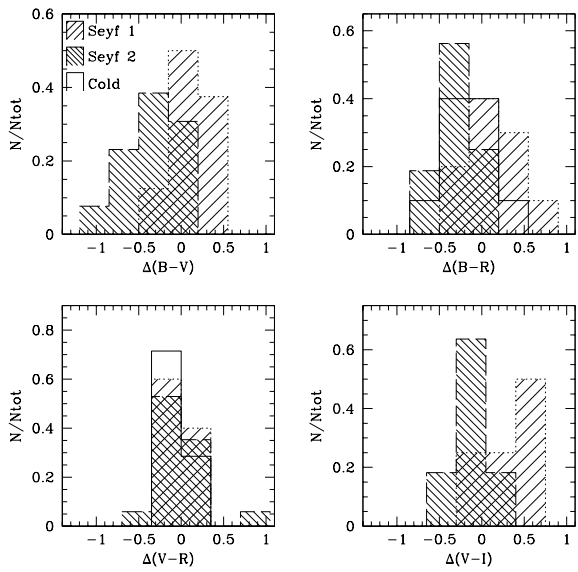}
\caption{Distributions of colour gradients, obtained from aperture (disk)-(nuclear) colours, for our three samples. On the $\Delta(B-V)$ and $\Delta(V-I)$ histograms the Cold sample is not shown, due to the small number of data points. \label{f1}}
\end{figure}

\begin{figure}
\epsscale{1.}
\plotone{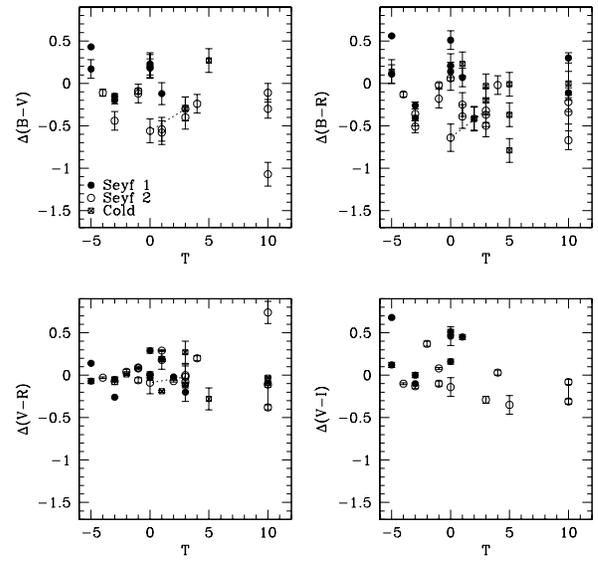}
\caption{Colour gradients obtained from aperture (disk)-(nuclear) colours versus morphological type T. T=10 is conventionally used for highly disturbed or double nucleus merger systems. Full lines connect members of an interacting system that are the possible IRAS candidates. Dashed lines connect the two nuclear colours of double nucleus merger systems. \label{f2}}
\end{figure}

We use the nuclear (within a 2 kpc radius) and disk (between 2 kpc and the 
isophote corresponding to $\mu_{B}$=25 mag arcsec$^{-2}$) aperture colours 
from Paper II, in which we have also presented the colour distributions and 
colour correlations for each sample. We define the aperture colour gradients simply 
as \[\Delta(Colour)=Colour_{(Disk)}-Colour_{(Nucleus)}\]. Thus, a 
positive/negative colour gradient corresponds to a redder/bluer disk
compared to the central 2 kpc region. In Figure~\ref{f1} we plot the
distributions of colour gradients and in Figure~\ref{f2} the gradients
as a function of host morphological type T, for our three (sub)samples.
The corresponding median, mean and standard deviations are given in 
Table~\ref{tab1}.

\placefigure{f1}

\placefigure{f2}

\placetable{tab1}

We summarize our main conclusions:

(i) Seyfert type 1 and 2 galaxies show opposite colour gradients (positive-negative, respectively).
This was already noticed by us in Paper II, as well as, from the 2D colour 
maps and the radial colour profiles presented in the 
Appendix of Paper III. The distribution of $(V-R)$ gradients is narrower 
compared to the other colours and peaks around zero. This is likely a combined
effect of (a) the short wavelength range, (b) the contamination of the 
$R$-band magnitudes (nuclear in the case of Seyfert 1, disk in the case of 
Seyfert 2 and Cold galaxies) by H$\alpha$ emission and (c) the contamination 
of the $V$-band magnitudes by [\ion{O}{3}]$_{4959,5007}$ emission. 
Seyfert 1s show positive gradients for all colours, $(B-V)$, $(B-R)$, $(V-I)$,
which are most likely due to the presence of a ``naked'' AGN in their
centers, that bluens significantly their nuclear colours (see also Figure 2 of
Paper II). The negative colour gradients in Seyfert 2s are comparable to those found 
in normal galaxies, for which stellar population age and metallicity effects 
are the main contributors (\cite{jong96c}). In Seyfert 2s, reddening of the central regions due
to dust obscuration could further steepen their 
colour gradients. Supporting evidence to the latter is the progressive flattening
(less negative) of the colour gradients towards longer wavelengths.  

(ii) The Cold sample galaxies all show negative outwards colour gradients. They are
less steep than the Seyfert 2 gradients and are comparable for the $(B-R)$ and
$(V-R)$ colours. These two facts are likely indicating that dust extinction 
and star formation are more evenly distributed throughout these objects.

(iii) There is no well-defined correlation between colour gradients and 
morphological type (with perhaps the exception of the Seyfert 2 $(V-I)$ 
gradients). The lack of correlation with morphological type for the Warm 
Seyferts, most likely
indicates that the aperture colour gradients are affected by both the AGN
and dust extinction, rather than simply reflecting stellar age and
metallicity gradients. Morphological missclassifications, due to the complex 
host morphologies, is another likely reason for the lack of correlations with 
T, in particular for the highly disturbed, interacting Cold sample galaxies.

\subsection{Surface Colour Gradients}

\subsubsection{Definitions}

\begin{figure}
\epsscale{1.}
\plotone{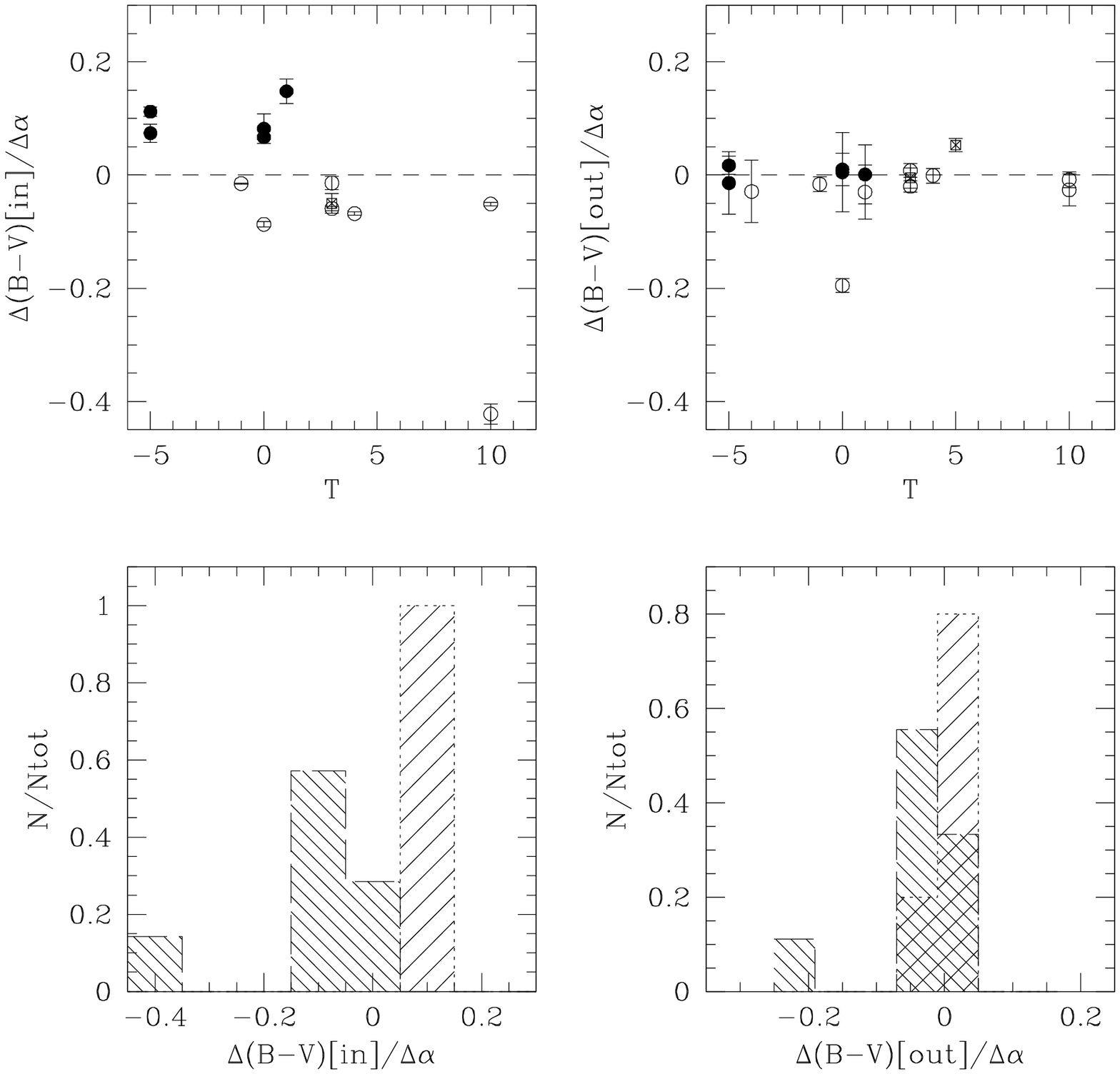}
\caption{Disk surface $\mu_{B}-\mu_{V}$ {\em inner} and {\em outer} colour gradients (obtained as described in the text). Upper panels: versus morphological type T (symbols as in Figure 2). Lower panels: distributions for the Seyfert 1 and 2 samples (symbols as in Figure 1).\label{f3}}
\end{figure}

\placefigure{f3}

\begin{figure}
\epsscale{1.}
\plotone{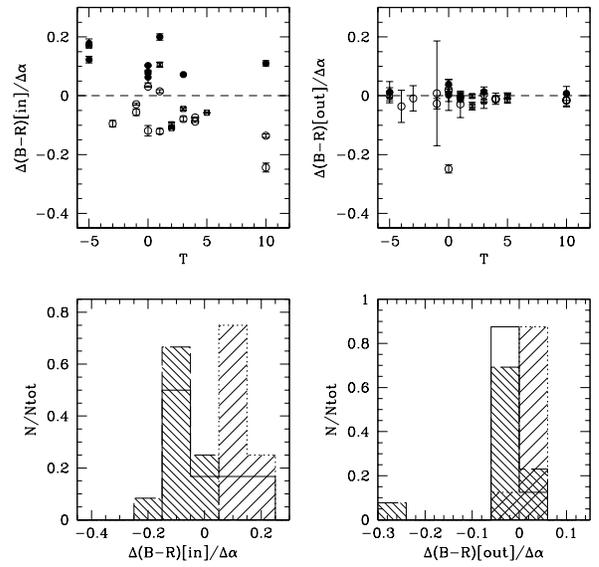}
\caption{Disk surface $\mu_{B}-\mu_{R}$ {\em inner} and {\em outer} colour gradients (obtained as described in the text). Upper panels: versus morphological type T (symbols as in Figure 2). Lower panels: distributions for the three (sub)samples (symbols as in Figure 1). (The Seyfert 2 galaxy that shows the atypically large error bar in its outer $(B-R)$ gradient is IRAS 03202-5150, due to its faint $B$ image). \label{f4}}
\end{figure}

\placefigure{f4}

\begin{figure}
\epsscale{1.}
\plotone{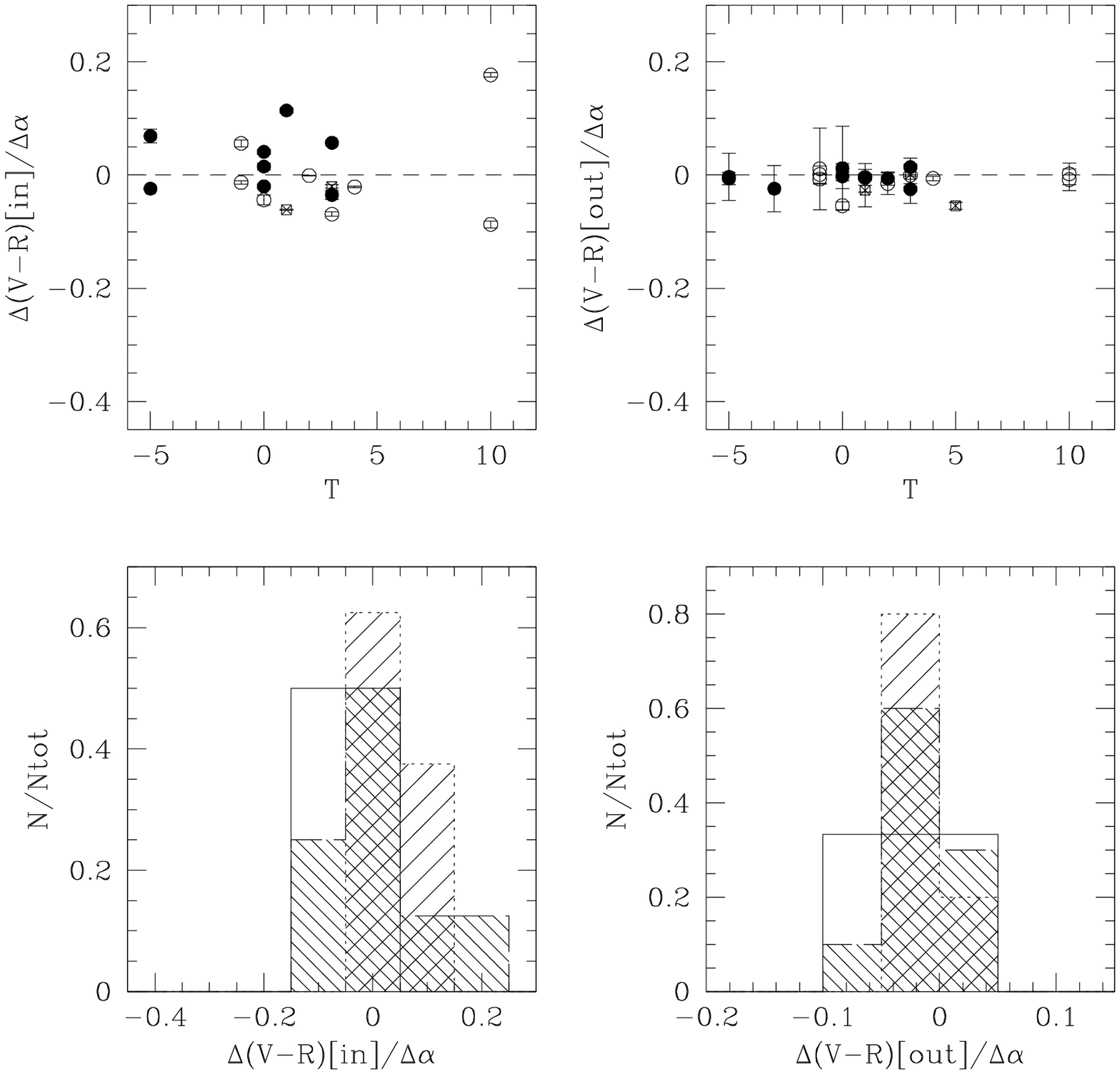}
\caption{Disk surface $\mu_{V}-\mu_{R}$ {\em inner} and {\em outer} colour gradients (obtained as described in the text). Upper panels: versus morphological type T (symbols as in Figure 2). Lower panels: distributions for the three (sub)samples (symbols as in Figure 1).\label{f5}}
\end{figure}

\placefigure{f5}

We used the isophotal fitting results from Paper III, to derive colours at 
each radius from the azimuthally averaged isophotal magnitudes. We are mainly 
interested in isolating the host colour gradients, thus we only use the 
data for radii larger than 2 kpc out to the $\mu_{B}$=25 mag arcsec$^{-2}$.
For elliptical galaxies, it is common to plot radial colour profiles versus 
log radius because these are essentially linear functions. However, this is 
not the case for our objects and thus we choose to represent colour gradients 
in linear radius space. We then derived the colour gradients 
\(\frac{\Delta(Colour)}{\Delta\alpha}\) from the slope of the radial colour 
profiles. Here $\alpha$ represents the semi-major axis length and {\em Colour}
the surface colour. In this way, we measured colour gradients through 
weighted least-square fitting of a first-order polynomial, using the surface 
photometry errors to weigh the data. Inspection of the colour profiles in the
Appendix of Paper III, shows that there is almost always a break in the 
profiles, particularly well-defined in the case of Seyfert 1s. We thus decided
to fit most colour profiles in two regions, {\em inner} and {\em outer} from 
the break location, which is defined to be the point where the inner
(linear) gradient stops fitting the data points. The outer portion of the 
profile is then fitted from the larger radii inwards. The gradients obtained 
in this
way, are presented in Table~\ref{tab2} (subscripts {\em I} and {\em O} for the
inner/outer gradients, respectively) and in Figures~\ref{f3} -~\ref{f5} 
(subscripts {\em in} and {\em out}). Defining the 
break point from the inner fits has the advantage that is less affected by 
the presence of structure in the galactic disks. The break radius (in kpc) is also listed
in Table~\ref{tab2}. When only one line is fitted to the totality of the 
profile, this is considered as {\em outer} disk gradient. The median, mean and
standard deviations of the colour gradient distributions are listed in
Table~\ref{tab3}. The uncertainties in the calculated gradients are comparable
for the different samples and for the various colours. The median errors are:
0.008 for \(\frac{\Delta(B-V)_{I}}{\Delta\alpha}\) and 
\(\frac{\Delta(B-R)_{I}}{\Delta\alpha}\), 0.004 for 
\(\frac{\Delta(V-R)_{I}}{\Delta\alpha}\), 0.017 for 
\(\frac{\Delta(B-V)_{O}}{\Delta\alpha}\) and 
\(\frac{\Delta(B-R)_{O}}{\Delta\alpha}\) and 0.019 for
\(\frac{\Delta(V-R)_{O}}{\Delta\alpha}\). The errors in the outer gradients are 
larger by a factor of $\sim$10, due to the presence of structure in the disks 
and to sky subtraction uncertainties. Comparison of the median errors quoted 
above and the values listed in Tables~\ref{tab2} and ~\ref{tab3}, indicates 
that the inner $(B-V)$ and $(B-R)$ gradients are well defined (with errors at 
the $\sim$10\% level or less), while the inner $(V-R)$ and all the outer 
gradients are very small (flat colour profiles), comparable to their 
associated errors.

\begin{table}
\dummytable\label{tab2}
\end{table}

\placetable{tab3}

\subsubsection{Results}

(i) All surface colour gradients are flatter (inner gradients are flatter by 
factors of 2 to 5) compared to the integrated ones, the former being less 
affected by nuclear light and thus more representative of the true disk 
gradients. This effect is more noticeable in Seyfert 2s.
However, there still remains a pronounced difference between the Seyfert 1 
and Seyfert 2 inner gradients, the former being always positive the 
latter mostly negative. If the persistent inverse gradients in Seyfert 1s are 
due to contamination by nuclear light, this would mean that the AGN dominates 
the light at least inside the central 5 kpc (median break radius) in these 
objects. However, as we have shown in Paper III, it is very unlikely that the 
bulge component (resulting from our profile decomposition) in Seyfert 1s is contaminated by the AGN light. The outer disk
gradients are distributed in a much narrower range around zero, but the 
distinction between the two Seyfert samples persists, the Seyfert 1s showing 
positive gradients even in the outer disk portion. Since no positive 
metallicity gradients are predicted by galaxy formation theories, these are 
likely to be age gradients, the younger stars being more centrally 
concentrated. The Student's T and the K-S tests show that the Seyfert 1 and 
2 subsamples have statistically different inner {\em and} outer $(B-R)$ colour
gradients (statistical significance better than 98\%). We have checked for 
possible inclination effects, but we find no correlation between surface 
colour gradients and ellipticity for any of our samples. This is important, in
particular for the Seyfert 2s, indicating that the host colour gradients are 
not dominated by orientation effects (and thus dust extinction). 

The Cold sample surface colour gradients are also flatter than their 
integrated gradients (factor of $\leq$2) and are mostly overlapping with the 
Seyfert 2 gradients (Figures~\ref{f4} and~\ref{f5}.
The similarity between these two samples, noticed earlier for  
their disk luminosities, sizes and colours (Paper II), is likely to
indicate that similar processes dominate the formation of their disks.

\begin{figure}
\epsscale{1.}
\plotone{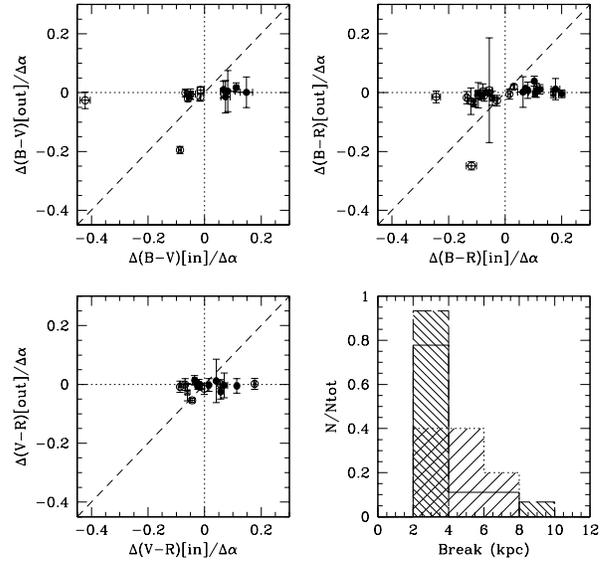}
\caption{{\em Inner} versus {\em outer} disk surface colour gradients (obtained as described in the text; symbols as in Figure 2). The long-dashed line indicates the line of equal colour gradients. The lower right panel shows the distribution of the radii where the break in colour profiles occur, for the three (sub)samples (symbols as in Figure 1).\label{f6}}
\end{figure}

(ii) No correlation was found between surface colour gradients and 
morphological type for any of our samples. If the profile break is indicative 
of bulge/disk dominance, there should be a correlation between break radius 
and morphological type, but we find no such obvious correlation.
On the contrary, for Seyfert 2 and Cold galaxies we find a very narrow range 
of break values (lower right panel of Figure~\ref{f6}), that is 
remarkably similar for the two (sub)samples, peaking around 3.5-4 kpc (with a 
dispersion of $\sim$1.5). The break radius distribution for Seyfert 1s is 
flatter and has a larger median $\sim$5 kpc. Modeling of the colour gradients 
of elliptical galaxies has shown that condensed dust distribution can produce 
a similar break in the colour profiles, marking the transition between 
optically thick and thin material. The location of this break provides then 
useful constraints on the extent and mass of the dust within a galaxy 
(\cite{wise96}). Our data show that this effect is likely to be more important
in the case of Seyfert 2 (and possibly Cold) galaxies.

(iii) Next, we investigated the change in slope at the break point. 
In Figure~\ref{f6} we plot inner versus outer surface colour 
gradients, the long-dashed line indicating the loci of equal gradients. Points
to the right/left side of this line indicate steeper (more positive/negative,
respectively) {\em inner} gradients. The furthest the points lie from this 
line the largest is the change in slope between the inner and outer parts of 
the colour profiles. First of all, we notice that there are no objects for 
which the {\em sign} of the slope changes between the inner and outer disk 
regions (upper left or lower right quarters of each 
plot). Seyfert 1s have typically steeper inner gradients (that is, more 
dramatic changes in slope at the break point) than Seyfert 2s. The detached
point representing a Seyfert 2 galaxy with the most negative $(B-V)$ and 
$(B-R)$ and positive $(V-R)$ inner gradients is the double nucleus merger 
IRAS 13536+1836; its extreme colour gradients are due to the light 
contamination by the two nuclei, at $\alpha$=2 kpc.
The other extreme Seyfert 2 galaxy, showing very blue outer
colour gradients is IRAS 11298+5313W, the western member of a triple interacting 
system and one of the possible candidates for the associated IRAS source. Its 
distinct position from the rest of the sample might be indicating that it
does not belong to the IRAS Warm sample (the E member of the triple system is 
the most probable IRAS candidate). 

\placefigure{f6}

\begin{figure}
\epsscale{1.}
\plotone{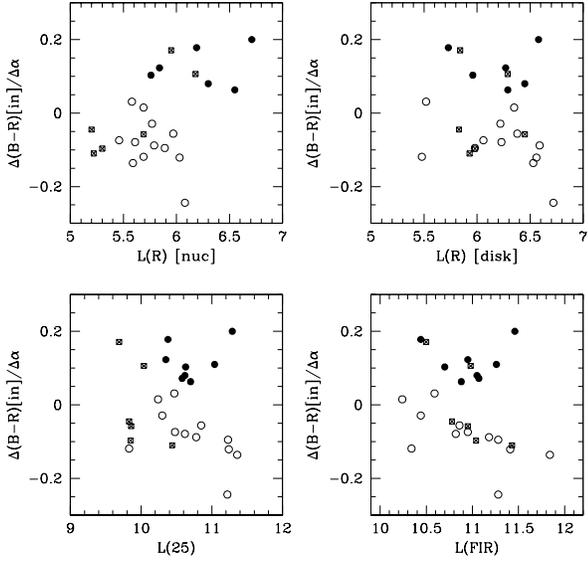}
\caption{Disk surface colour gradients versus optical and IR luminosities (in units of log(\Lsun); the errors associated with the data points were omitted here for clarity; symbols are as in Figure 2).\label{f7}}
\end{figure}

(iv) We now address the question of whether any correlations exist between the
colour gradients and the optical or IR luminosities. There is a well-known 
difference in colour gradients (excluding dust effects) between  bright and 
faint bulges of ellipticals and early type galaxies, that are clues to
different formation mechanisms for these galaxies (\cite{balcells94}). In fact, 
this dichotomy in colour gradients with absolute magnitude seems to persist 
even outside their central regions (\cite{vader98}). In Figure~\ref{f7}
we show plots of colour gradients versus luminosities, for our data. We find no correlation between 
optical or IR luminosities and colour gradients for Seyfert 1s, but we do see 
a correlation between inner $(B-R)$ gradients and IR luminosities (both
$L_{25}$ and $L_{FIR}$) for Seyfert 2s, in the sense of bluer gradients at 
larger IR luminosities. In fact, these correlations become very tight if we
exclude the two points deviating towards steeper gradients, that represent the
galaxies IRAS 13536+1836 and 11298+5313W mentioned earlier to have unusually 
steep gradients compared to the rest of the sample. A Spearman Rank or a 
Kendall's Tau non-parametric test for the significance of the correlations give
0.02 for \(\frac{\Delta(B-R)_{I}}{\Delta\alpha}\) vs $L_{25}$ and 0.007 for 
\(\frac{\Delta(B-R)_{I}}{\Delta\alpha}\) vs $L_{FIR}$. A similar trend is
seen, for the inner $(B-R)$ gradients to correlate with $L_{FIR}$ for the
Cold sample, but the data points are fewer and the scatter larger (correlation
significance 0.04). 

There are two possible ways to interpret the tight correlation seen for 
Seyfert 2s, given that the IR luminosity scales with the amount of (warm) dust
within a galaxy: (a) if the dust is concentrated in the central regions, we
expect negative (outwards) colour gradients that are steeper the larger is the
effect of dust reddening in the center (b) if the warm dust is distributed 
throughout the disk, larger IR luminosities indicate a 
stronger source of illumination, most probably strong disk star formation 
that also causes the optical colours to bluen outwards. If alternative (a) is
correct,
then we would expect that a fair amount of dust heating in the center would 
be due to the AGN and thus the above correlation would be better defined 
versus $L_{25}$. Alternative (b) on the other hand, would be better 
represented by the correlation versus $L_{FIR}$. In fact, since both
correlations are observed, the most probable explanation is a combination of 
the two effects. 

(v) To seek an explanation for the above correlations, we inspected the 
continuum images (Paper I) and colour maps (Appendix of Paper III) for our 
Seyfert 2 sample. We find a clear correlation between increasing interaction 
strength and larger IR luminosities or steeper colour gradients. This result 
can be understood if strong interactions bring larger amounts of gas and dust 
to the center, induce strong star formation events throughout the disk and in 
the central regions and maybe feed the AGN, thus increasing the level of 
nuclear activity. \cite{bushouse90} have also noticed larger scale and steeper
optical and near-IR colour gradients in interacting and star-forming galaxies,
compared to normal spirals. This result will be explored in more detail in 
Paper V.

\placefigure{f7}

\subsubsection{Discussion}

Having explored the colour distributions and radial gradients for our sample
galaxies, we compare now our data to normal galaxies and other Seyfert 
samples. Elliptical and early type galaxies are evolved systems, almost 
entirely composed by old red stars. Colour gradients in these galaxies are 
always negative outwards and can mostly be explained by metallicity effects. 
This is 
probably also the case for the bulges of early type spirals, although their 
colours are in general bluer than ellipticals, indicating that the bulge 
stellar populations are younger and/or more metal poor (\cite{balcells94}). 
Moreover, their colour gradients do not seem to correlate with galaxy type 
(which is also what we found for the inner gradients of our sample galaxies). 
Negative colour gradients are also typical for intermediate and late type 
spirals. Here, many other effects are present besides metallicity gradients: 
variable internal extinction, dust re-radiation and the distribution of 
different stellar population types, can all influence the observed (bulge and
disk) colours and gradients (\eg \cite{jong99,jong96c}). It is clear that the 
behaviour of our Warm 
Seyfert 2 and Cold galaxies is closer to that observed for normal spirals, 
the former having steeper colour gradients than the latter, indicating that 
dust effects in their central regions are important. 
Although the integrated colour gradients of Seyfert 1s are influenced by the 
presence of the AGN in their centers, it is not clear why the inverse 
(positive outwards) gradients persist also outside the central 2 kpc region 
(surface colour gradients). Positive gradients, when found in dwarf ellipticals,
are interpreted as stellar age gradients: the evolution in these objects is
dominated by internal processes, such as galactic winds and dissipation, that 
prevent the formation of metallicity gradients in these objects (\cite{vader98}). 

The differing colour properties of the disks of Warm Seyfert 1 and 2 galaxies 
are likely to be intrinsic and thus difficult to reconcile with a simple 
orientation effect. There is little information in the recent literature about
Seyfert colour gradients. \cite{kotilainen94} computed optical and near-IR 
{\em aperture} colour gradients for a sample of hard X-ray selected, mainly 
Seyfert 1 galaxies. After subtracting the AGN contributions they find negative
optical and IR gradients, comparable to those of normal galaxies and certainly
smaller than those found in interacting and starburst galaxies. They 
give $\Delta(B-V)$=-0.07 which is comparable to the inner surface colour 
gradients that we found for our Seyfert 2 sample. In fact, \cite{kenty90} has 
found redder disks than nuclei for his sample of Markarian and NGC Seyferts, 
in particular for objects with amorphous or peculiar morphologies (see also 
Figure~\ref{f9}). The range of his colours and medians are comparable to
our results for the IR Warm Seyferts. We computed the mean {\em integrated} 
(Disk-Nucleus) colour gradients for MacKenty's data and find mean 
$(B-V)$=0.14, $(B-R)$=0.09, $(V-R)$:-0.05. These are comparable with our data 
for the Warm Seyfert 1 aperture colour gradients (Table~\ref{tab1}). However,
MacKenty's definition of nuclear and total aperture colours is different
than ours, which makes this comparison less meaningful. 

It is difficult to quantitatively interpret our results, given the
multiplicity of factors that can affect the colours within a galaxy. Colour 
gradients that are primarily due to dust extinction are positive inwards at 
any given wavelength, but their morphology and magnitude will depend on the 
amount and distribution of the dust. Assuming some given dust properties, the 
colour gradients should then have similar overall morphologies at all 
wavelengths, but the slopes will be much steeper at shorter wavelengths 
(larger optical depths). Moreover, because these slopes are likely to be 
strongly affected by stellar population (age and metallicity) gradients as 
well, the latter effects are better explored in the near-IR where colour 
gradients are more insensitive to dust. Without such near-IR data it is 
impossible to discriminate between stellar population and dust as sources of 
the observed optical colour gradients. Even with near-IR data, metallicity and
extinction could still be degenerate and other types of data, such as high 
resolution spectra and/or resolved images redwards of the near-IR, are 
necessary to settle the question unambiguously.

\section{Stellar Populations}

Many systematic studies of star formation rates (SFR) in spiral galaxies
show that the variation of optical colours and H$\alpha$ properties through 
the Hubble sequence is due to different birthrate histories in the galactic 
disks: early type (S0-Sb) galaxies formed most of their stars in less than 
t$_{Hubble}$, but late type systems (Sc-Irr) form stars in a constant rate 
since their birth and will continue to form stars for several Gyr 
(\eg \cite{kennicutt94} and references therein). Modeling of the broad band 
colours is a very useful tool in this respect, providing that one uses disk 
colours that are unaffected by the old spheroidal component, rather than 
integrated colours. This approach was used by a number of workers, in order to study 
normal galaxies. \cite{kennicutt94} and \cite{devereux91}
suggested that changes in photometric properties along the Hubble sequence are
purely due to the evolutionary (star formation) history of disks and nearly 
independent of the changing $B/D$ ratio. \cite{kennicutt94} found that the 
observed properties of disks involve an initial mass function (IMF) which is 
enriched in massive stars by factors 2-3 over the solar neighborhood IMFs of 
\cite{miller79} and \cite{scalo86}. He also found that a finite-time recycling
of the gas (returned from stars or/and through interactions) increases the 
life-time of the gas up to 5-15 Gyr (compared to 3 Gyr for instantaneous 
recycling). An important conclusion reached by Kennicutt was that the 
disk SFR per unit luminosity changes dramatically through the Hubble sequence 
(0.01-0.1 in Sa-Sb, 0.5-2 in Sc), while this change is much smaller within 
individual disks. The opposite however was found to be true for the (mean and 
radial respectively) disk surface densities. These two findings suggest that 
what determines the SFR in galactic disks is not merely the local gas surface 
density. \cite{jong96c} found for his sample of face-on spirals, that their 
radial colour gradients are matched by differences in the star formation 
{\em histories} (SFH) within a galaxy. The outer parts are populated by young 
stars, while the red colours of the central regions require a range of 
metallicities in a relatively old stellar population. \cite{jong99} have 
reached a different conclusion than \cite{kennicutt94}, suggesting that, in
spiral galaxies, the local surface density is the most important parameter in 
determining the star formation history and together with the galaxy mass, 
the galaxy's chemical evolution.

In this section we are going to investigate the stellar population content of 
our sample galaxies, based on their IR properties and our optical photometry 
results.

\subsection{Star Formation Rates}

Optical and IR luminosities measure star formation over slightly different 
time periods: the H$\alpha$ emission is mainly due to ionization by young stars
($\leq$10$^{7}$yr) while the IR luminosity is a measure of star formation over
the past 10$^{8}$-10$^{9}$yr. We have calculated approximate SFRs for our 
objects, based on their far-IR ($\geq$60 $\mu$m) luminosities. We use the 
standard formula
\(SFR_{IR}=1.3\times 10^{-43}L_{FIR}\Msun yr^{-1}\) ($L_{FIR}$ in ergs
cm$^{-2}$ sec$^{-1}$), adopted from \cite{hunter86}, for a Salpeter IMF 
(\cite{salpeter55}) between 0.1-100 \Msun. Although IR luminous galaxies seem 
to require a larger IMF (\eg \cite{rieke80,rieke93}), this is more complicated
to assess than in the case of normal galaxies, because of the heavy 
obscuration in those systems and the highly time-variable SFRs. The 
conventional IMF chosen here is sufficient to make an inter-comparison of our 
samples.

\begin{figure}
\epsscale{0.8}
\plotone{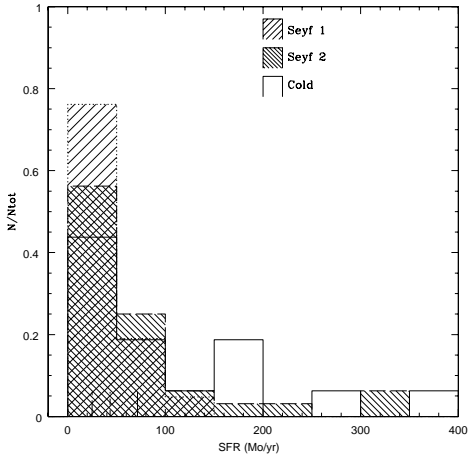}
\caption{Distribution of star formation rates obtained from far-IR emission for our samples. The vertical bars on the lower x-axis indicate median values for each sample: long-dashed bar for Seyfert 1s, short-dashed for Seyfert 2s and full-line bar for the Cold sample.\label{f8}}
\end{figure}

In Figure~\ref{f8} we plot the histograms of SFR$_{IR}$ for our three 
subsamples and their median values as vertical bars. All samples span a large 
range of SFRs, in particular the Cold sample (9-380 \Msun yr$^{-1}$) which has
a larger median 71.5 \Msun yr$^{-1}$ than the Warm sample. The Seyfert 1 and Seyfert 2 subsamples
span similar ranges ($\sim$5-150) with medians 24.9 and 38.6 \Msun yr$^{-1}$ 
respectively, excluding the four highly IR luminous ($L_{60}\geq$10$^{11}$) 
Seyfert 2s that have SFR$_{IR}\geq$150 \Msun yr$^{-1}$. When the four 
ultraluminous (UL) objects are included the two samples have significantly
different variances and means (significance 0.0001 and 0.02, respectively). 

We should be cautious when interpreting the FIR luminosities as reliable
indicator of the star formation activity in galaxies. It is important to 
compare samples for which the (precursor) galaxy dust properties are known and
similar. For instance, there is very likely an intrinsic difference in the 
dust content of galaxies selected at 60 $\mu$m, compared to galaxies selected 
optically. Also, bigger galaxies reradiate more energy at all wavelengths
(ideally one should normalize $L_{FIR}$ by the galaxy size).
Moreover, $L_{FIR}$ is due to different dust components, only the warmer
component being associated with star formation. However, in dusty star forming
regions with large optical depth, the stellar radiation field is dominated by
young stellar populations and thus $L_{FIR}$ effectively measures the 
$L_{bol}$ of the starburst. Moreover, in Paper II we have shown that the 
$L_{FIR}$ for our samples is related to the disk component, thus it is less dependent on the AGN.
Consequently, the assumption of $L_{FIR}$ being an accurate star formation indicator
seems to be a good approximation for the objects in our (IR-selected) samples.
Keeping in mind the limitations outlined above, it is instructive to compare
our results with those of some other characteristic galaxy samples.
Using the equivalent widths of H$\alpha$ emission, \cite{kennicutt83} 
found a SFR (extinction corrected) as high as 20 \Msun yr$^{-1}$ in giant Sc 
galaxies and suggested that in late-type spirals, current SFRs are similar to 
past SFRs averaged over the age of the disk. SFRs obtained from IR data are 
usually larger than those from optical emission; \cite{armus90} argued that a 
factor of $\sim$3 difference between the two values is within the 
uncertainties attached to the models used for the calculation of SFRs and to 
the poorly known amount of extinction that affects the H$\alpha$ emission. 
For a sample of isolated galaxies, drawn from the complete IRAS bright 
galaxy sample (with $f_{60}\geq$5.24 Jy) the mean SFR$_{IR}$ is 9.9 \Msun
yr$^{-1}$ (\cite{soifer89}). A sample of Markarian on-going and advanced 
merging systems (most of them starbursts and some Seyferts) shows median 
SFR$_{H\alpha}$=6.8 \Msun yr$^{-1}$ and SFR$_{IR}$=12.5 \Msun yr$^{-1}$
(\cite{mazzarella91}). In these multiple-nucleus systems, warm dust is 
concentrated in the nuclear regions and is heated by active star formation, 
producing the bulk of far-IR emission and their warm far-IR colours. It is 
clear that our IR-warm Seyfert samples show larger values than their Markarian
(UV-selected) counterparts.
A far-IR warm (colour-selected) sample of powerful FIR galaxies 
(\cite{armus90}) shows mean SFR$_{H\alpha}$=40 \Msun yr$^{-1}$ and 
SFR$_{IR}$=117 \Msun yr$^{-1}$. These objects are suspected to be recent 
mergers undergoing strong circumnuclear bursts of star formation that ionize 
the interstellar medium throughout the galaxy. Our Cold sample
was not selected according to far-IR warmness as the Armus \etal sample, but 
it is directly comparable to it in terms of $L_{FIR}$ and 
$\alpha_{(25,60)}$.
The ULFIRG sample ($L_{FIR}\geq$10$^{12}$\Lsun) shows an even larger mean 
SFR$_{IR}$=313 \Msun yr$^{-1}$. This is a sample of extremely bright, strongly
interacting systems, their far-IR emission being associated with large scale 
star formation, triggered by the interaction.

\cite{mazzarella91} argued that the differences in IR properties (and SFRs) 
between these samples can be convincingly explained by invoking differing 
relative fractions and temperatures of the warm dust, increasing from 55\% and
T$_{d}$=40 K for the Markarian sample to 80\% and T$_{d}$=50 K for the ULFIRGs
sample. They suggested that there are possible correlations between 
{\em strong} encounters and increased far-IR emission, enhanced star formation
and/or nuclear activity. This agrees with the conclusions that we reached
earlier from our colour gradient data (see previous section) and is also 
consistent with the SFRs that we find for our sample objects: the Cold 
galaxies (mostly strongly interacting systems) show larger values by factors 
2-3 compared to the Warm Seyfert sample. The four UL Warm Seyfert 2s with 
large IR luminosities and SFRs are also merging/interacting systems. The issue
of interactions versus galactic activity will be further explored in Paper V.

\placefigure{f8}

\subsection{Colour-Colour Profiles}

Colour-colour profiles have been commonly used in order to model the
effects of dust and stellar populations (age-metallicity) in the colours of
galaxies. \cite{larson78} have shown that, while normal galaxies follow a 
well-defined relation ($(U-B)$ vs $(B-V)$ in their case), peculiar and 
interacting galaxies show a much larger scatter, with overall bluer colours. 
This difference was attributed to the latter having experienced anomalous
SFRs, characterized by recent bursts. Several attempts to model galactic 
colours were put forward since then, the most sophisticated being stellar 
population synthesis models. Generally, there are several trends appearing in 
colour-colour diagrams: (a) The dispersion along the normal sequence indicates
different rates of star formation, the colours becoming bluer for increasing 
SFR. However, metallicity (chemical composition) changes within a galaxy or 
among galaxies will have a similar effect on colours, that is, will move the 
colours parallel to the normal sequence, redder for higher metallicities. 
(b) The dispersion perpendicular to the normal sequence indicates differences 
in star formation bursts, the more recent (or shorter) bursts corresponding to
bluer colours. However, another similar effect is produced by changing IMF, 
bluer colours being produced by more massive stars (flatter IMF). The 
degeneracy between these various factors is a major problem for the modeling
of galaxy colours and gradients. Another major complication, is the degeneracy
between stellar population and dust effects. IR data is very important for 
disentangling these effects and a variety of extinction models exist in the 
literature. \cite{jong96c} computed extinction models for his sample of 
face-on spiral galaxies and found that the dust changes neither the shape of 
the disk light profiles (in his case exponential) or the colour gradients: 
varying the internal extinction produces only colour and surface brightness 
$\mu$ {\em offsets} (see also \cite{jong99}). Similarly, for a sample of 
edge-on galaxies, 
\cite{just96} found that dust effects being constant in the optically thick 
central regions, do not affect the intrinsic colour gradients. They just 
{\em shift} their colours. If this is true in all galaxies it alleviates 
somewhat the degeneracy problem, providing that one is not interested in the
absolute galaxy colours. 

\subsubsection{Integrated Colours}

\begin{figure}
\epsscale{1.}
\plotfiddle{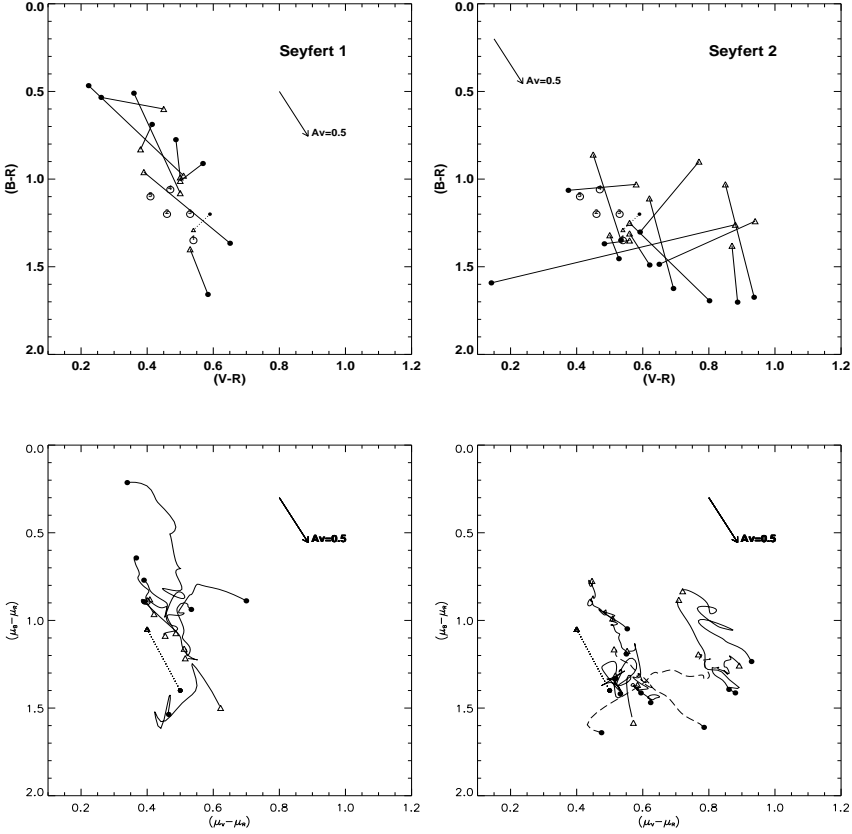}{400pt}{0}{60}{50}{-150}{-20}
\figcaption{Upper panels: Aperture $(B-R)$ colour-colour plots; filled circles are nuclear and triangles disk colours. The arrow indicates the Galactic extinction law. The circles indicate the loci of {\em integrated} colours for a sample of face-on normal spirals (1:T=0-2, 2:T=2-4, 3:T=4-6, 4:T=6-8. 5:T=8-10; De Jong 1996).The dashed line and smaller symbols indicate average colours for the Seyfert sample of MacKenty 1990. Lower panels: Surface $(B-R)$ colour-colour plots, left panel Seyfert 1s, right panel Seyfert 2s; filled circles are colours at 2 kpc from the nucleus and triangles correspond to colours at the $\mu_{B}$=23 mag arcsec$^{-1}$ isophote. The dashed line indicates an average profile for normal spirals (above reference) and the long-dashed lines in the Seyfert 2 plot indicate the tracks of double nucleus merger systems.\label{f9}}
\end{figure}

A large amount of exctinction must be present in all our sample galaxies. 
However, since we have no good measure for the internal exctinction and we are
mainly interested in comparing the shapes (rather than the absolute values) of
colour gradients between our objects, we shall not consider the effects of 
extinction in the discussion that follows.
In Figure~\ref{f9} we show $(B-R)$ vs $(V-R)$ plots for the Warm Seyfert
1 and 2 subsamples. The upper panels show integrated nuclear (filled dots) and
disk (triangles) colours. The arrow indicates the Galactic extinction law for
$A_{V}$=0.5 that, although an oversimplification of the real dust 
distribution, it is a gross indicator of the direction of dust exctinction 
effects (\cite{jong96c}). The nuclear and disk symbols are connected together 
to indicate aperture colour gradients (discussed in Section 3.1). Overplotted 
circles indicate the loci of integrated colours 
for a sample of face-on normal spirals (\cite{jong96c}), with morphological 
types T=0-10. The dashed line connecting smaller symbols represents the mean
nuclear and disk colours for a sample of Markarian and NGC Seyfert galaxies 
(\cite{kenty90}), which were found to possess in a large extent disturbed 
morphologies.

The first thing depicted here is the difference in the {\em sign} of the 
colour gradients between Seyfert 1 and 2 types, which was discussed earlier. 
The only two Seyfert 1s with a negative gradient (bluer 
disks) are faint objects (IRAS 04124-0803 and 13512-3731) and thus their 
location in this plot should be regarded with caution.
Another immediate observation is that Seyfert 1s and 2s lie in different parts
of the diagrams, the former having bluer colours than normal galaxies the 
latter mostly redder, this including both nuclear and disk (and thus 
integrated) colours. When computing differences between the mean 
{\em integrated} colours for our three subsamples and for the De Jong sample 
of normal spirals, we find that for all colours, except $(B-V)$, Seyfert 1s 
are bluer by $\sim$0.1-0.3 mag while Seyfert 2s (and Cold galaxies) redder by 
$\sim$0.05-0.2 mag. All our samples have $(B-V)$ colours that are
bluer than normal galaxies. Although the magnitude of these differences is 
comparable to the colour errors given by \cite{jong96c}, they are consistent 
in sign for all the galaxies within our sample, indicating that the effect 
is real. The most likely interpretation is that the bluer colours of Seyfert 
1s are due to contamination by the AGN, while the bluer $(B-V)$ 
colours in Seyfert 2s are due to strong star formation in their {\em disks} 
(their nuclear colours being redder than normal galaxy colours).  A third 
observation is that the {\em nuclear} colours for each of the two Seyfert 
samples appear to lie along some type of sequence in Figure~\ref{f9}. We
do not find this to be a sequence of Hubble types, interaction stage or of 
some other morphological peculiarity. In Paper II we have seen that the Warm
Seyfert 1 nuclear colours scale with \(\frac{L_{FIR}}{L_B}\). This was 
interpreted in terms of dust obscuration, which is likely to be represented by
the sequence in the colour-colour plots. The trend for Seyfert 2 nuclei is 
more difficult to understand. There is one Seyfert 2 galaxy with apparently 
very blue $(V-R)$ colour; this is IRAS 13536+1836, a double nucleus system for
which only the E nucleus is considered here (see also Paper II).

\placefigure{f9}

\subsubsection{Surface Colours}

Because integrated colours are usually dominated by the nucleus,
a better approach is to use radial colour profiles, especially for 
the study of disk populations. In the lower panels of Figure~\ref{f9} we
plot the $(\mu_{B}-\mu_{R})$ vs $(\mu_{V}-\mu_{R})$ tracks starting from 
radius 2 kpc (filled circles) out to $\mu_{B}$=23 mag arcsec$^{-1}$ 
(triangles). The long-dashed lines on the Seyfert 2 plot indicate the double 
nucleus mergers IRAS 13536+1836 and 19254-7245. Not all objects plotted on the
upper panels appear on the lower (those with shallow images were omitted). The
distribution of points is similar between the upper and lower panels, but the 
tracks are quite different and will be discussed below in terms of comparison 
with population synthesis models. The dotted lines represent an average track 
for the normal spiral sample of \cite{jong96c} for 0$\leq$T$\leq$5, that is 
shifted towards bluer $(V-R)$ colours compared to both our Seyfert samples. 
For Seyfert 1s, the outer disk $(B-R)$ colours are quite similar to the spiral
disk colours, while the inner $(B-R)$ colours are much bluer than the spiral 
nuclear colours. Seyfert 2s on the other hand have similar inner and outer 
disk $(B-R)$ colours with those of normal spirals, but significantly redder 
$(V-R)$ colours. Moreover, the Seyfert 2 tracks are often complex, due to 
additional components affecting their light profiles. De Jong has shown that 
all his galaxies have a negative gradient outwards that can be best explained 
by a combination of stellar age and metallicity effects, the outer parts being
populated by younger and lower metallicity stars while the dust plays a minor 
role to the observed gradients. He also found a correlation between colours 
and surface brightness, both within and among galaxies, and a correlation 
between colours and morphological type T for the same surface brightness
(see also \cite{jong99}). As 
we have shown in Paper III, there is no significant correlation between 
morphological type and surface brightness (or any other disk parameter) in our
samples. Neither do we find any correlations between colours or colour 
gradients and morphological type. 

\subsection*{\em Comparison with models}

\begin{figure}
\epsscale{1.}
\plotfiddle{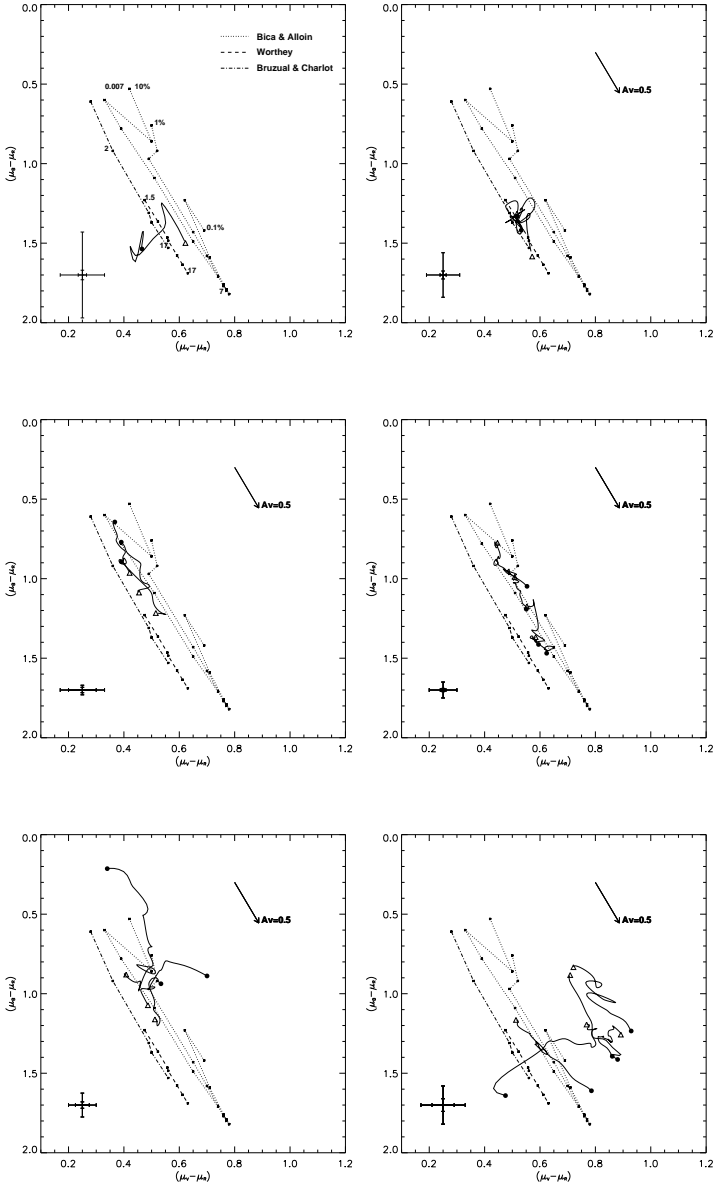}{400pt}{0}{60}{60}{-150}{0}
\caption{Surface $(V-R)$ vs $(B-R)$ colour-colour plots (full lines): left panels for Seyfert 1s right panels for Seyfert 2s. Filled circles are colours at 2 kpc from the nucleus and triangles correspond to colours at the $\mu_{B}$=23 mag arcsec$^{-1}$ isophote. Population synthesis models are overplotted, the coding given in the upper left panel. The small dots within each model represent stellar ages with the younger and older labeled in units of Gyr (the full range is given in the text). The \% labels in the three Bica \& Alloin models represent burst-to-old population mass fractions. The arrows indicate a standard Galactic extinction law. Error bars plotted on the lower left corner indicate the average inner (smaller) and outer (bigger) errors for the objects in that panel.\label{f10}}
\end{figure}

In view of the age-metallicity-extinction degeneracy discussed in the previous
section, we shall not attempt here to give an individual comparison between 
our data and stellar population synthesis models. Instead, we shall discuss 
the main trends for our sample objects using as reference some well-known 
population synthesis models, chosen to include both age and metallicity 
effects.
In Figure~\ref{f10} we split each sample into three (somewhat arbitrary)
groups whose colour-colour tracks show some similarities. The left panels 
are reserved for Seyfert 1s, the right panels for Seyfert 2s. The population 
synthesis models are overplotted in each panel and decoded on the top left 
panel. Here is a short description of these models:

(i) \cite{bica90} (dotted line): Models of star forming events of the last 
3$\times$10$^{9}$ yr are superimposed on an old stellar populations. Here, 
young, low-metallicity star clusters are combined with the spectra of red 
strong-lined nuclei, to simulate the effects of starbursts induced by galaxy 
interactions, either through gas inflow from the disk to the nucleus of a 
spiral galaxy or through gas transfer from a spiral disk or irregular system
to an earlier type galaxy. A library of observed star cluster spectra is 
used with IMF similar to those in starbursts. Three different models, labeled 
by 0.1\%, 1\% and 10\%, are shown in Figure~\ref{f10}, representing the 
respective burst-to-old population mass fractions. The dots in all three 
models represent burst ages: $\leq$0.007, 0.02, 0.07, 0.2, 0.7, 2, 7 Gyr. 

(ii) \cite{worthey94} (dashed line): These models are calculated for a wide 
range of metallicities, but represent only a range of intermediate to old 
stellar populations: 1.5, 2, 3, 5, 8, 12, 17 Gyr. These are single-burst
models using the standard Salpeter IMF with 0.1$\leq$M$\leq$2 \Msun. In 
Figure~\ref{f10} we show only the model corresponding to solar 
metallicity, mainly for comparison purposes with the other models presented 
here and because different metallicity tracks are almost superimposed on each 
other for our plotted scales.

(iii) \cite{bruzual93} (dash-dotted line; see also \cite{charlot91}): Stellar 
population synthesis models, calculated for solar metallicities and starting 
from the very early stages of evolution: $\leq$0.0001, 2, 7, 12, 17 Gyr. These
models use a Salpeter IMF with 0.1$\leq$M$\leq$100 {\Msun} and are calculated 
for different star formation histories, that is, a single burst, an 
exponentially declining SFR and a constant SFR model. On our plots we chose to
represent the single burst model, for compatibility with the other models and 
also because this should be the most likely situation in Seyferts.

The Bruzual \& Charlot and Worthey models are offset from each other (although
the relative trends are similar), this being due to uncertainties in the used
stellar evolutionary tracks (\cite{jong96c}). It was indeed found that 
large discrepancies exist between different population synthesis models, the 
uncertainties in stellar age for a given metallicity can be as large as 35\% 
and in metallicity for a given age of the order of 25\% (\cite{charlot96}).
That is, uncertainties in these two quantities are larger than any effects 
from changing IMFs or cut-offs. The main effects of model uncertainties are 
shifts in absolute colour values, but the predicted colour gradients and other
relative trends within a galaxy or between different galaxies should be 
correct and directly comparable with the observations (see \eg \cite{jong96c}).

\placefigure{f10}

Let us now briefly discuss the main conclusions drawn from 
Figure~\ref{f10}. Throughout this discussion, the reader should be
referring to the detailed colour profiles and two-dimensional maps shown in
the Appendix of Paper III.

First of all, it is clear that our galaxies have very different profiles than 
those of normal spirals (shown as dotted line in Figure~\ref{f9}), which
were better fitted by a combination of the Worthey and Bruzual \& Charlot 
models (\cite{jong96c}).

(a) Upper panels: The Seyfert 1 galaxy on the left panel is IRAS 13512-3731 
and the two Seyfert 2s on the right panel are IRAS 04507+0358 and 03202-5150. 
The first is a compact early-type galaxy; its nuclear blue colours are 
probably affected by the AGN and slightly reddened outwards through a track 
that is difficult to interpret given the large error bars. This object is 
redder than the rest of the Seyfert 1 sample; according to the Worthey or 
Charlot \& Bruzual models, the indicated mean stellar ages are $\sim$5-8
Gyr. The first of the two Seyfert 2s mentioned above, IRAS 04507+0358, is also
an early type galaxy whose colours lie within the above two sequences of 
models, indicating somewhat younger mean populations (within the model and 
data uncertainties) than the Seyfert 1 galaxy. The second Seyfert 2 galaxy,
IRAS 03202-5150, is an early type galaxy, most likely interacting with a 
nearby companion. The main body has colours indicative of a relatively old 
stellar population (a mean of 5 Gyr according to the Worthey models). This 
galaxy has also a bar and ring features with associated star formation (see
Appendix of Paper III), that are seen as loops in its colour profile and as a
blue jump towards the loci of (star forming) models of \cite{bica90}.

(b) Middle panels: The Seyfert 1 galaxies on the left panel are IRAS
02366-3101, 04493-6441 and 15015+1037. They are all isolated, early or 
intermediate type galaxies, with progressively redder colours in the cited 
order. They all have inverse (positive) gradients and their tracks are 
parallel and partially overlap with the Bica \& Alloin tracks, indicating a 
very recent starburst (0.2-0.7 Gyr if the 10\% model is to be used) superposed
on an older stellar population. Given the size of the error bars, the colours 
could also fit the 1\% model and a $\leq$0.1 Gyr starburst or, less likely,  a
Bruzual \& Charlot model, with mean stellar ages $\sim$5 Gyr or younger.

The Seyfert 2 galaxies on the right panel are IRAS 03059-2309, 23254+0830, 
02580-1136 and 20481-5715, this being a range of progressively redder mean 
colours. The latter is an early type galaxy with a nearby companion. There is 
little change in its colours, which show a 1-2 Gyr stellar population 
superposed on the redder underlying galaxy. IRAS 02580-1136 is an early-type 
barred spiral with strong star formation associated with its ``grand design'' 
spiral arms. Its colours are consistent with a young 10\% starburst $\sim$1-2 
Gyr old (or 0.1-0.2 Gyr if the 1\% model is used), from disk to center. 
The other two Seyfert 2 galaxies are both later type spirals and members of 
strongly interacting systems. Both show grand design spiral arms (and tidal 
tails) with knotty star forming regions. Their colour profiles overlap with 
the 1\% and 10\% Bica \etal models, indicating recent star formation: 0.5-1
Gyr for the 10\% model, or younger by a factor of 10 for the 1\% model 
in IRAS 23254+0830 and 0.2-0.7 Gyr or younger in IRAS 03059-2309 (the range
of stellar ages is from disk to center).

In summary, all the objects presented in these two panels have colour profiles
consistent with the evolutionary tracks of Bica \etal, but inconsistent with 
either of the other two models shown in Figure~\ref{f10}. Changes in 
metallicity would have only a minor effect (for instance may be responsible 
for the slightly differing slopes of the profiles), as it is apparent from
the virtually overlapping Worthey models for a wide range of metallicities.
The Charlot \& Bruzual models for different star formation {\em histories} 
(not shown in Figure~\ref{f10}) do not resemble our colour profiles and 
lie far from our data points. Although the influence of the AGN makes it more 
difficult to interpret the Seyfert 1 inverse colour gradients, it is likely
at least for the Seyfert 2s, that their colour gradients mainly represent 
stellar age effects, if extinction was to be ignored (dust extinction would
have a similar effect on the colour profiles). Although for a simple Galactic 
exctinction law such an effect could not be much larger than $\sim$0.5 mag in 
$V$ band {\em within} a galaxy, different assumptions for the dust 
distribution could result in larger effects (\eg \cite{jong96c} and references
therein).

(c) Lower panels: In these plots we show objects with complex and unusual 
colour profiles that cannot be explained easily by a simple comparison with 
population models. The Seyfert 1s plotted in the left panel are IRAS
00509+1225, 23016+2221 and 21299+095. The first one (bluest of the three) is 
an early type spiral with knotty emission in its (one-sided) spiral arm. What 
makes it unusual is its very blue nuclear region. Its disk colours could be 
fitted with a very young starburst $\leq$0.5 Gyr old, superposed on the old 
galaxy population. IRAS 23016+2221 (with the bluest $(V-R)$ nuclear colour) is
an object with peculiar morphology, reminiscent of a recent merger, originally
given an early Hubble classification. Its strong H$\alpha$ emission within the
central $\sim$5 kpc (see Appendix of Paper III) is probably responsible for 
the bluer $(V-R)$ disk compared to its nucleus. Modeling its colours with 
Bica \& Alloin's tracks indicates a very young starburst of mean age 0.07-0.7 
Gyr. The last object is IRAS 21299+095, an early type spiral which also shows 
strong circumnuclear H$\alpha$ emission, that causes its mean disk colours to 
become significantly bluer (compared to the nucleus). They can be fitted by a 
starburst of mean age 0.02-0.5 Gyr. It is difficult to explain its very red 
$(V-R)$ nuclear colour, though. 
On the right panel we show some extreme Seyfert 2 cases: IRAS 13144+4508,
11298+5313 (W and E members), 19254-7245 and 13536+1836 (in order of bluening
$(V-R)$ nuclear colours). The first three objects have similar $(B-R)$ colours
as the rest of the Seyfert 2 sample and their colour profiles are (roughly) 
following the evolutionary tracks of Bica \& Alloin. However, their $(V-R)$ 
colours are strongly shifted to the red by $\sim$0.2-0.3 mag, distinguishing 
them from the rest of the sample. They are all members of multiple strongly 
interacting systems, suffering tidal distortions and they all have very blue 
$(B-V)$ (integrated) disk colours, indicating recent star formation events.
They fit the description given by \cite{larson78} of strongly interacting 
galaxies (at a particular stage of their dynamical evolution, involving tidal 
features), showing very blue $(B-V)$ colours, off the normal sequence in
colour-colour plots. The latter two objects are examples of on-going mergers,
with double nuclei embedded in a common body and large tidal tails extending 
radially outwards. Actually, the colours of IRAS 19254-7245 can be explained 
with a Bica \& Alloin type of model, indicating starbursts with age
$\sim$0.1-5 Gyr or younger if we allow for the (certainly very significant) 
dust effects, invoked in order to explain the steep colour gradient
and the very red nuclear colour. On the other hand, it is very difficult
to explain the colour profile of IRAS 13536+1836, part of the problem being 
the very blue W nucleus that affects the mean nuclear colour at 2 kpc.

We have not presented colour-colour profiles for any of the Cold sample 
objects. The only two objects among them with three-colour information are 
IRAS 07514+5327 and 06506+5025, both late-type barred spirals. On 
colour-colour diagrams as the ones of Figure~\ref{f10}, their colour 
profiles are similar to those of the third group Seyfert 2 galaxies (lower 
panels). They show very red colours that cannot be interpreted by any of the 
population synthesis models and their profiles vary almost perpendicular to 
the normal colour-colour sequence. These effects seem to be independent of 
the nuclear activity type (one object is a starburst, the other a Seyfert 1
galaxy).

We summarize our conclusions in the next section.

\section{Conclusions}

In this paper we have examined the colour distributions characterizing the
host galaxies of our Warm and Cold samples. Our main conclusions are as 
follows:

1. The Warm Seyfert 1 galaxies show bluer nuclei and positive colour 
gradients, while the Warm Seyfert 2 and the Cold galaxies show the exact 
opposite trends. Most likely, these aperture gradients reflect the 
contamination of nuclear colours by the AGN in Seyfert 1s and by dust 
extinction in Seyfert 2s. In the Warm Seyfert 2 and Cold samples, the colour 
and emission line distributions are more complex than in the Warm Seyfert 1s, 
showing evidence for patchy dust extinction and intense star formation mostly 
associated with spiral and tidal features.

2. Surface colour gradients at radii $\geq$2 kpc, indicate a clear distinction
between the Warm Seyfert 1 and 2 galaxies: For the Seyfert 1s, positive colour
gradients persist even at large disk radii ($\geq$5 kpc) where contamination 
by the nucleus is minimal, most likely indicative of (increasing outwards) age
gradients. In Seyfert 2s, any nuclear exctinction should not affect seriously 
the colour gradients at radii$\geq$2 kpc, which are thus dominated by 
(decreasing outwards) metallicity and stellar age effects. There is an overall
similarity between Warm Seyfert 2 and Cold galaxy colour gradients, which 
indicates that similar (external) processes, such as strong interactions, must
dominate their disk properties.

3. Seyfert 2 (bluer) colour gradients correlate with (larger) IR luminosities,
indicating centrally concentrated dust and strong disk star formation. In 
fact, we find that both these observed quantities scale with interaction 
strength, a result that will be further explored in Paper IV.

4. The Cold sample shows larger SFRs (as deduced from their IR emission 
longword of 60 $\mu$m) by factors of 2-3 compared to the Warm Seyfert samples.
However, if in the Seyfert 2 subsample we include the ultra-luminous 
($L_{60}\geq$10$^{11}$) (strongly interacting) members, it becomes 
statistically comparable to the Cold sample. Galactic interactions seem again 
to be the issue here, for the enhanced far-IR emission and SFRs.

5. The radial colour-colour profiles show that the Warm Seyfert 1 and Seyfert 
2 galaxies occupy different regions (the first bluer, the second redder) in 
these diagrams. For each sample the galaxies can be grouped in three classes: 
(a) Early-type galaxies with mean disk stellar ages $\geq$5 Gyr (b) Early or 
intermediate-type galaxies with colours indicating starbursts $\sim$1 Gyr or
younger, superposed on the older underlying galaxy population. Within this 
class, the Seyfert 1s are isolated objects while the Seyfert 2s are mostly 
interacting with a companion galaxy. (c) Objects with complex profiles and,
in the case of Seyfert 2s, also complex morphologies. The Seyfert 1s are 
mostly single early-type systems with strong circumnuclear H$\alpha$ emission 
and colours indicating a very young starburst $\leq$0.5 Gyr superposed on the 
redder galaxy population. The Seyfert 2s in this class, are all extreme cases 
of strongly interacting, tidally distorted and double nucleus merging systems,
that lie way off the normal sequence, with extremely red $(V-R)$ colours. 
The only two Cold galaxies with available three-colour information, belong also
to this latter class of extreme profiles. We conclude that, for all three 
classes, the observed colour profiles can be described by age and dust effects
within single-burst, solar metallicity models. Metallicity changes within or 
among the galaxies and differing star formation histories do not seem to 
affect significantly the above conclusions.

6. In double nucleus systems one of the two nuclei is activated and becomes 
the main source of optical and IR emission, ionizing the gas anisotropically. 
These characteristics are strikingly similar for the four mergers that we have
observed, independently of which sample they belong to. 

In the present Paper IV we have shown significant differences in the host
Seyfert 1 and 2 colour distributions, that cannot be attributed to simple 
orientation effects. In turn, they seem to be related to the interaction stage
of the host galaxy, at least for the Warm Seyfert 2 galaxies which in this
respect are similar to the Cold galaxies. Through aperture photometry
(Paper II) and decomposition of the host light profiles (Paper III) we have 
reached similar conclusions. In the forthcoming, last in this series, Paper IV
we will combine all these results within the context of an evolutionary 
scenario for the Warm Seyfert galaxies.

\acknowledgments
I am grateful to my thesis advisors George Miley and Walter Jaffe for providing
me with stimulation and support throughout the completion of this project.
This research has made use of the NASA/IPAC Extragalactic Database (NED) 
which is operated by the Jet Propulsion Laboratory, California Institute of 
Technology, under contract with the National Aeronautics and Space 
Administration. Part of this work was completed while the author held a 
National Research Council - NASA GSFC Research Associateship.

%
%

\clearpage

\begin{deluxetable}{lrrrrrr}
\tablenum{1}
\tablecolumns{7}
\tablewidth{0pc}
\tablecaption{Median, Mean and Dispersion of the Aperture Colour Gradient Distributions. \label{tab1}}
\tablehead{
\colhead{} & \colhead{Seyf 1} & \colhead{Seyf 2} & \colhead{Cold} & \colhead{Seyf 1} & \colhead{Seyf 2} & \colhead{Cold} \\
\cline{1-7}
 & & & & & & \nl
\colhead{Quantity} & \multicolumn{3}{c}{$(B-V)_{disk-nucleus}$} & \multicolumn{3}{c}{$(B-R)_{disk-nucleus}$} \\
}
\startdata
	Median & 0.175 & -0.300 & -0.015\tablenotemark{*} & 0.125 & -0.330 & -0.115 \nl
	Mean   & 0.092 & -0.374 & -0.015\tablenotemark{*} & 0.112 & -0.304 & -0.186 \nl
	$\sigma$ & 0.223 & 0.275 & 0.310\tablenotemark{*} & 0.310 & 0.215 & 0.311 \nl
 & & & & & & \nl
\cline{1-7} \\
Quantity & \multicolumn{3}{c}{$(V-R)_{disk-nucleus}$} & \multicolumn{3}{c}{$(V-I)_{disk-nucleus}$} \nl
 & & & & & & \nl
\cline{1-7} \\
	Median & -0.025 & -0.020 & -0.090 & 0.305 & -0.100 & \nodata \nl
	Mean   & 0. & 0.042 & -0.060 & 0.285 & -0.093 & \nodata \nl
	$\sigma$ & 0.169 & 0.234 & 0.175 & 0.277 & 0.204  & \nodata \nl
\enddata
\tablenotetext{*}{Few data points}
\end{deluxetable}

\clearpage

\begin{deluxetable}{lrrrrrr}
\tablenum{3}
\tablecolumns{7}
\tablewidth{0pc}
\tablecaption{Median, Mean and Dispersion of the Surface Colour Gradient Distributions. \label{tab3}}
\tablehead{
\colhead{} & \colhead{Seyf 1} & \colhead{Seyf 2} & \colhead{Cold} & \colhead{Seyf 1} & \colhead{Seyf 2} & \colhead{Cold} \\
\cline{1-7}
 & & & & & & \nl
\colhead{Quantity} & \multicolumn{3}{c}{\(\frac{\Delta(B-V)_{I}}{\Delta\alpha}\)} & \multicolumn{3}{c}{\(\frac{\Delta(B-V)_{O}}{\Delta\alpha}\)} \\
}
\startdata
	Median & 0.082 & -0.059 & \nodata\tablenotemark{*} & 0.005 & -0.020 & 0.024\tablenotemark{*} \nl
	Mean   & 0.097 & -0.102 & \nodata\tablenotemark{*} & 0.004 & -0.035 & 0.024\tablenotemark{*} \nl
	$\sigma$ & 0.033 & 0.143 & \nodata\tablenotemark{*} & 0.011 & 0.061 & 0.041\tablenotemark{*} \nl
 & & & & & & \nl
\cline{1-7} \\
\colhead{Quantity} & \multicolumn{3}{c}{\(\frac{\Delta(B-R)_{I}}{\Delta\alpha}\)} & \multicolumn{3}{c}{\(\frac{\Delta(B-R)_{O}}{\Delta\alpha}\)} \\
 & & & & & & \nl
\cline{1-7} \\
	Median & 0.110 & -0.079 & -0.045 & 0.011 & -0.012 & -0.002 \nl
	Mean   & 0.116 & -0.083 & -0.005 & 0.011 & -0.029 & -0.010 \nl
	$\sigma$ & 0.050 & 0.073 & 0.114 & 0.013 & 0.068  & 0.013  \nl
 & & & & & & \nl
\cline{1-7} \\
\colhead{Quantity} & \multicolumn{3}{c}{\(\frac{\Delta(V-R)_{I}}{\Delta\alpha}\)} & \multicolumn{3}{c}{\(\frac{\Delta(V-R)_{O}}{\Delta\alpha}\)} \\
 & & & & & & \nl
\cline{1-7} \\
	Median & 0.041 & -0.001 & -0.020\tablenotemark{*} & -0.003 & -0.004 & -0.027\tablenotemark{*} \nl
	Mean   & 0.027 & 0.021  & -0.405\tablenotemark{*} & -0.005 & -0.008 & -0.027\tablenotemark{*} \nl
	$\sigma$ & 0.052 & 0.081 & 0.028\tablenotemark{*} &  0.013 &  0.017 &  0.028\tablenotemark{*} \nl
 & & & & & & \nl
\cutinhead{ Break (kpc) }
 & \multicolumn{2}{c}{Seyfert 1} & \multicolumn{2}{c}{Seyfert 2} & \multicolumn{2}{c}{Cold} \nl
\cline{2-7}
 Median & \multicolumn{2}{c}{5} & \multicolumn{2}{c}{4} & \multicolumn{2}{c}{3.5} \nl
Mean & \multicolumn{2}{c}{4.9} & \multicolumn{2}{c}{4.3} & \multicolumn{2}{c}{4} \nl
$\sigma$ & \multicolumn{2}{c}{1.5} & \multicolumn{2}{c}{1.6} & \multicolumn{2}{c}{1.6} \nl
\enddata
\tablenotetext{*}{Few data points}
\end{deluxetable}

%
%

\clearpage

\clearpage

\clearpage

\clearpage

\end{document}